\documentclass[aps,prb,twocolumn,superscriptaddress,showpacs]{revtex4-1}

\usepackage{graphicx}
\usepackage{calc}
\usepackage{bm}
\usepackage{color}
\usepackage{textcomp}

\begin{document}

\title{Multiple magnon modes  in the  Co$_3$Sn$_2$S$_2$ Weyl semimetal candidate}

\author{O.O.~Shvetsov}
\author{V.D.~Esin}
\author{A.V.~Timonina}
\author{N.N.~Kolesnikov}
\author{E.V.~Deviatov}
\affiliation{Institute of Solid State Physics of the Russian Academy of Sciences, Chernogolovka, Moscow District, 2 Academician Ossipyan str., 142432 Russia}

\date{\today}

\begin{abstract}
 We experimentally investigate electron transport in kagome-lattice ferromagnet Co$_3$Sn$_2$S$_2$, which is regarded as a time-reversal symmetry broken Weyl semimetal candidate. We demonstrate $dV/dI(I)$ curves  with  pronounced asymmetric  $dV/dI$ spikes, similar to those attributed to current-induced spin-wave excitations in ferromagnetic multilayers. In contrast to multilayers, we observe several  $dV/dI$ spikes' sequences at low, $\approx$10$^4$ A/cm$^2$, current densities for a thick single-crystal Co$_3$Sn$_2$S$_2$ flake in the regime of fully spin-polarized bulk.  The spikes at low current densities can be attributed to novel magnon branches in magnetic  Weyl semimetals, which  are predicted due to the coupling between two magnetic moments mediated by Weyl fermions. Presence of spin-transfer effects at low current densities in Co$_3$Sn$_2$S$_2$ makes the material attractive for applications in spintronics.
\end{abstract}


\maketitle

\section{Introduction}
    
    A strong area of interest in condensed matter physics is topological semimetals~\cite{armitage}. Dirac semimetals  host the 4-fold degenerate topologically protected Dirac points, which are the special points of Brillouin zone with three dimensional linear dispersion. In Weyl semimetals (WSM), by breaking time reversal or inversion symmetries, the 4-fold degeneration of Dirac points declines to the 2-fold one, so every Dirac point splits  into two Weyl nodes with opposite chiralities.   Similarly to topological insulators and quantum Hall insulators, Weyl semimetals have topologically protected  surface states, which are Fermi arcs connecting projections of Weyl nodes on the surface Brillouin zone~\cite{armitage}. First experimentally investigated WSMs  were non-centrosymmetric crystals with broken inversion symmetry.  Spin- and angle- resolved photoemission spectroscopy data indeed demonstrate  spin-polarized surface Fermi arcs~\cite{das16,feng2016}. 

Also, ferromagnetic and antiferromagnetic WSMs with broken time-reversal symmetry can be discussed~\cite{armitage}. There are only a few candidates of magnetically ordered materials for the realization of WSMs~\cite{mag1,mag2,mag3,mag4}. Recently,  giant anomalous Hall effect was reported~\cite{kagome,kagome1} for the kagome-lattice ferromagnet Co$_3$Sn$_2$S$_2$, as an indication for the existence of a magnetic Weyl phase. Fermi arcs were also visualized~\cite{kagome_arcs} for Co$_3$Sn$_2$S$_2$  by scanning tunneling spectroscopy.

\begin{figure}
    \includegraphics[width=\columnwidth]{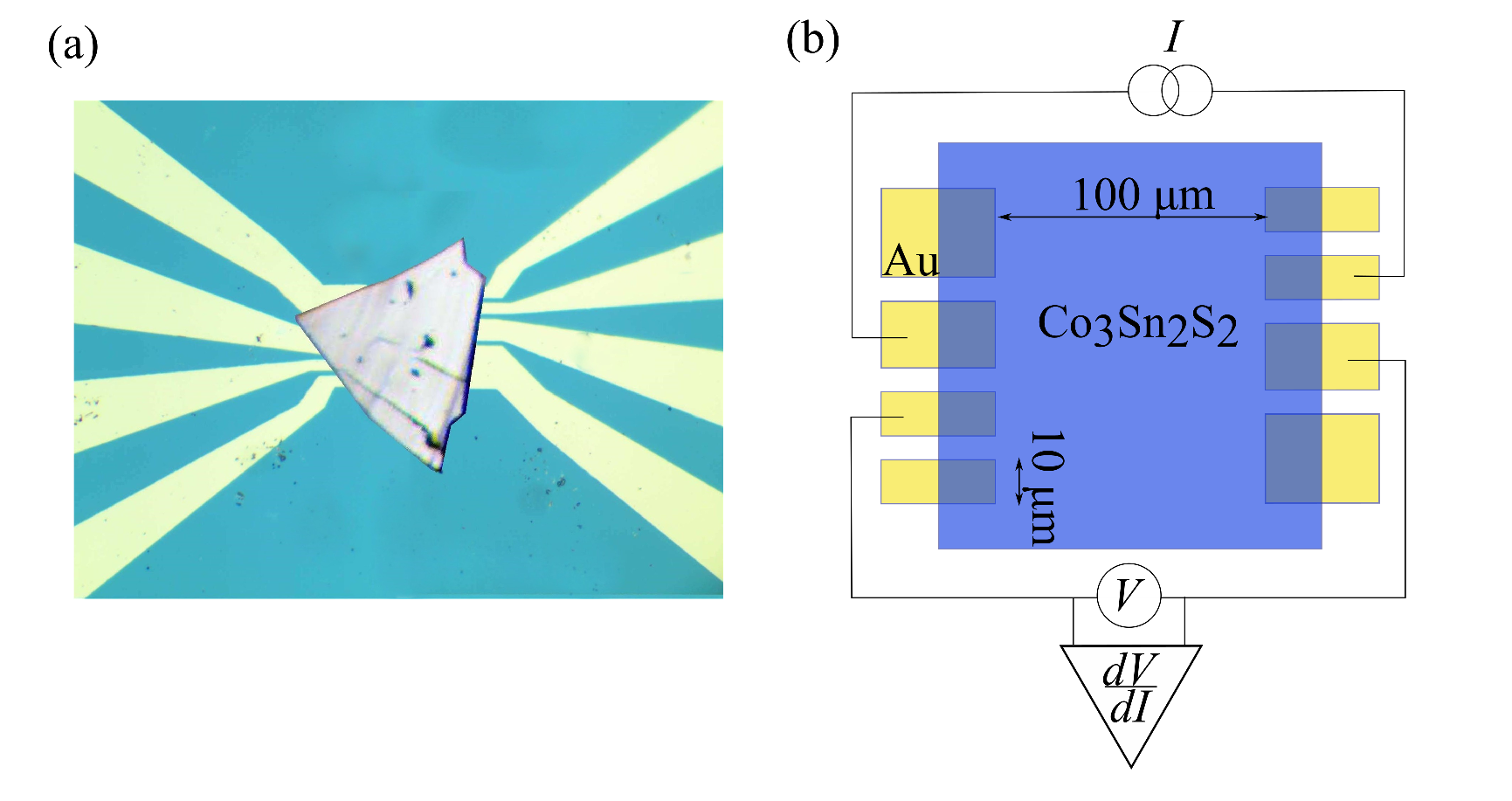}
    \caption{(Color online) (a) A top-view image of the sample. A flat (about 100~$\mu$m size and 1~$\mu$m thick) single-crystal Co$_3$Sn$_2$S$_2$ flake is weakly pressed on the insulating SiO$_2$ substrate with 100~nm thick, 5~$\mu$m separated  gold leads. The leads are of 40 $\mu$m, 20~$\mu$m, 10~$\mu$m and 10~$\mu$m  width  from up to down for the left side of the sample.  (b) The sketch of a sample with electrical connections. Non-linear $dV/dI(I)$ curves are measured by a standard four-point technique, so  a large   dc current $I$ is additionally modulated by a low   ac component. }
    \label{cosns_sample}
\end{figure}

Another specifics of time reversal symmetry broken WSMs is  a large anomalous Hall conductivity~\cite{kagome,kagome1}.
The anomalous Hall effect (AHE) manifests itself as non-zero Hall conductance in zero magnetic field. AHE is known in a large class of magnetically ordered materials due to the two qualitatively different microscopic mechanisms: extrinsic processes due to scattering effects, and an intrinsic mechanism connected to the Berry curvature~\cite{ahe1,ahe2,ahe3}. The latter variant is realized for  WSMs,  where the Berry curvature is enhanced at  Weyl nodes~\cite{armitage}. 

It is well known, that the magnetically ordered materials allows spin-wave excitations. For example, current-induced excitation of spin waves, or magnons,  was demonstrated in ferromagnetic multilayers at large electrical current densities~\cite{myers,tsoi1,tsoi2,katine,single,balkashin,balashov}. Spin-wave excitations are observed only for one current polarity and is presented as a sharp peak in differential resistance~\cite{myers,tsoi1,tsoi2,katine,single,balkashin,balashov}.  Electric field assisted magnetization dynamics was also predicted~\cite{weyl_magnon,dirac_magnon} for WSMs, which requires experimental investigations. In particular, novel magnon branches are predicted~\cite{weyl_magnon} in magnetic  Weyl semimetals, which  can be understood as a result of the coupling between two magnetic moments mediated by Weyl fermions.

Here, we experimentally investigate electron transport in kagome-lattice ferromagnet Co$_3$Sn$_2$S$_2$, which is regarded as a time-reversal symmetry broken Weyl semimetal candidate. We demonstrate $dV/dI(I)$ curves  with  pronounced asymmetric  $dV/dI$ spikes, similar to those attributed to current-induced spin-wave excitations in ferromagnetic multilayers. In contrast to multilayers, we observe several  $dV/dI$ spikes' sequences at low, $\approx$10$^4$ A/cm$^2$, current densities for a thick single-crystal Co$_3$Sn$_2$S$_2$ flake in the regime of fully spin-polarized bulk.  The spikes at low current densities can be attributed to novel magnon branches in magnetic  Weyl semimetals, which  are predicted due to the coupling between two magnetic moments mediated by Weyl fermions. Presence of spin-transfer effects at low current densities in Co$_3$Sn$_2$S$_2$ makes the material attractive for applications in spintronics.

\section{Samples and technique}

\begin{figure}
    \includegraphics[width=\columnwidth]{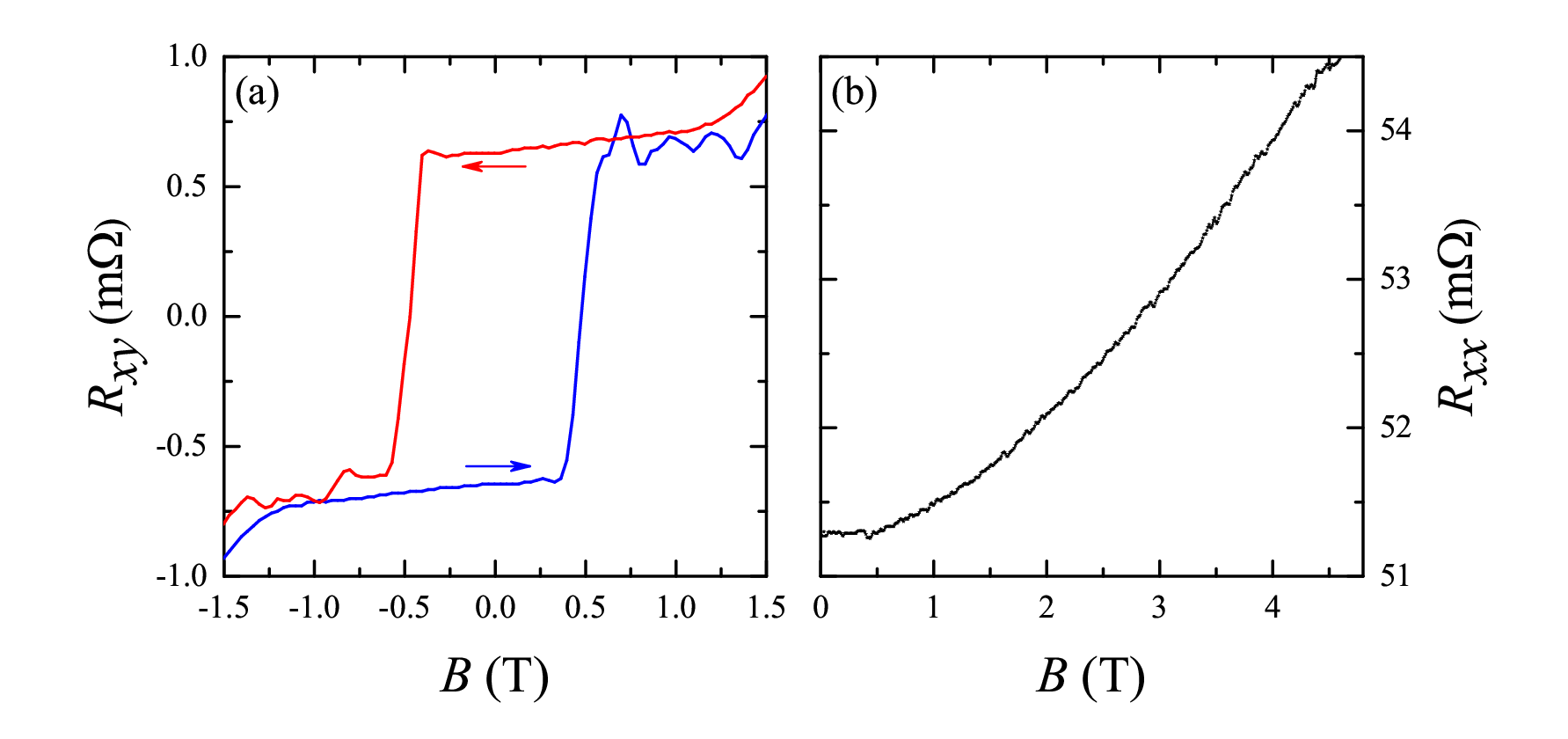}
    \caption{(Color online) (a) Giant anomalous Hall effect, which confirms~\cite{kagome} high quality of our kagome-lattice semimetal Co$_3$Sn$_2$S$_2$ samples~\cite{kagome,kagome1}. The Hall resistance $R_{xy}$ demonstrates  hysteresis behavior and sharp switchings at  $\approx 0.5$~T. Arrows indicate the scanning directions. (b) Positive, non-saturating longitudinal magnetoresistance  in normal magnetic field, which is a hallmark of compensated semimetals~\cite{kagome,kagome1}. The measurements are performed at 4.2~K. }
    \label{hall}
\end{figure}
   
    Co$_3$Sn$_2$S$_2$ single crystals were grown by the gradient freezing method. Initial load of high-purity elements taken in stoichiometric ratio was slowly heated up to 920$^\circ$~C in the horizontally positioned evacuated silica ampoule, held for 20 h and then cooled with the furnace to the ambient temperature at the rate of 20 deg/h. The obtained ingot was cleaved in the middle part. The Laue patterns confirm the hexagonal structure with $(0001)$ as cleavage plane.  Electron probe microanalysis of cleaved surfaces and X-ray diffractometry of powdered samples confirmed stoichiometric composition of the crystal.

Despite it is possible to form contacts directly on the  cleaved Co$_3$Sn$_2$S$_2$ crystal plane, large samples are not suitable for transport experiments: it is impossible to create high current density in a thick sample, which is crucial for the current-induced magnetization experiments.

Instead, the leads pattern is formed on the insulating SiO$_2$ substrate by lift-off technique after thermal evaporation of 100~nm Au. The gold leads are separated by 5~$\mu$m intervals, see Fig.~\ref{cosns_sample}~(a), and~(b). Since the kagome-lattice ferromagnet Co$_3$Sn$_2$S$_2$ can be easily cleaved along (0001) crystal plane, small (about 100~$\mu$m size and 1~$\mu$m thick) Co$_3$Sn$_2$S$_2$ flakes are obtained by a mechanical cleaving method. Then we select the most plane-parallel flakes  with clean surface, where no  surface defects could be resolved with optical microscope. They are transferred to the Au leads pattern and pressed slightly with another oxidized silicon substrate. A special metallic frame allows us to keep the substrates parallel and apply a weak pressure to the sample. No external pressure is needed for a Co$_3$Sn$_2$S$_2$ flake to hold on to a substrate with Au leads afterward.  This procedure provides reliable Ohmic contacts, stable in different cooling cycles, which has been also demonstrated before~\cite{cdas,nbwte,inwte}.

We check by standard magnetoresistance measurements that our Co$_3$Sn$_2$S$_2$ flakes  demonstrate giant anomalous Hall effect~\cite{ahe1,ahe2,ahe3} in normal to the flake's plane magnetic field. Fig.~\ref{hall} (a) shows hysteresis behavior and sharp switchings in Hall resistance $R_{xy}$, the switchings' positions $\approx 0.5$~T even quantitatively coincide with the previously reported ones~\cite{kagome,kagome1}.  Fig.~\ref{hall} (b)  shows positive, non-saturating longitudinal magnetoresistance, which is also consistent with the reported behavior~\cite{kagome} for Co$_3$Sn$_2$S$_2$ semimetal. Thus,  magnetoresistance measurements confirms high quality of our  Co$_3$Sn$_2$S$_2$ samples.

    We study electron transport along the Co$_3$Sn$_2$S$_2$ surface by a standard four-point technique. The principal circuit diagram is depicted in Fig.~\ref{cosns_sample}~(b).  To obtain $dV/dI(I)$ characteristics, a large (up to 3~mA) dc current $I$ is additionally modulated by a low ($\approx$5~$\mu$A) ac component. We measure both dc ($V$) and ac ($\sim dV/dI$) components of the voltage drop in Fig.~\ref{cosns_sample}~(b)  with a dc voltmeter and a lock-in, respectively, after a broad-band preamplifier. The lock-in signal is checked to be independent of the modulation frequency. The obtained $dV/dI(I)$ curves are qualitatively independent on the particular choice of current and voltage probes.  We check by comparison of four- and three-point $dV/dI(I)$ curves, that contact resistance is neglegible, see also Refs~\cite{cdas,nbwte,inwte} for details.

The measurements are performed in a usual He4 cryostat equipped with a superconducting solenoid.  Co$_3$Sn$_2$S$_2$ magnetic properties arise from the kagome-lattice cobalt planes, whose magnetic moments order ferromagnetically~\cite{kagome} out of plane below 175~K. We do not see noticeable temperature dependence in the interval 1.4-4.2~K, so all the results below are obtained at 4.2~K.

\section{Experimental results}

\begin{figure}
\includegraphics[width=\columnwidth]{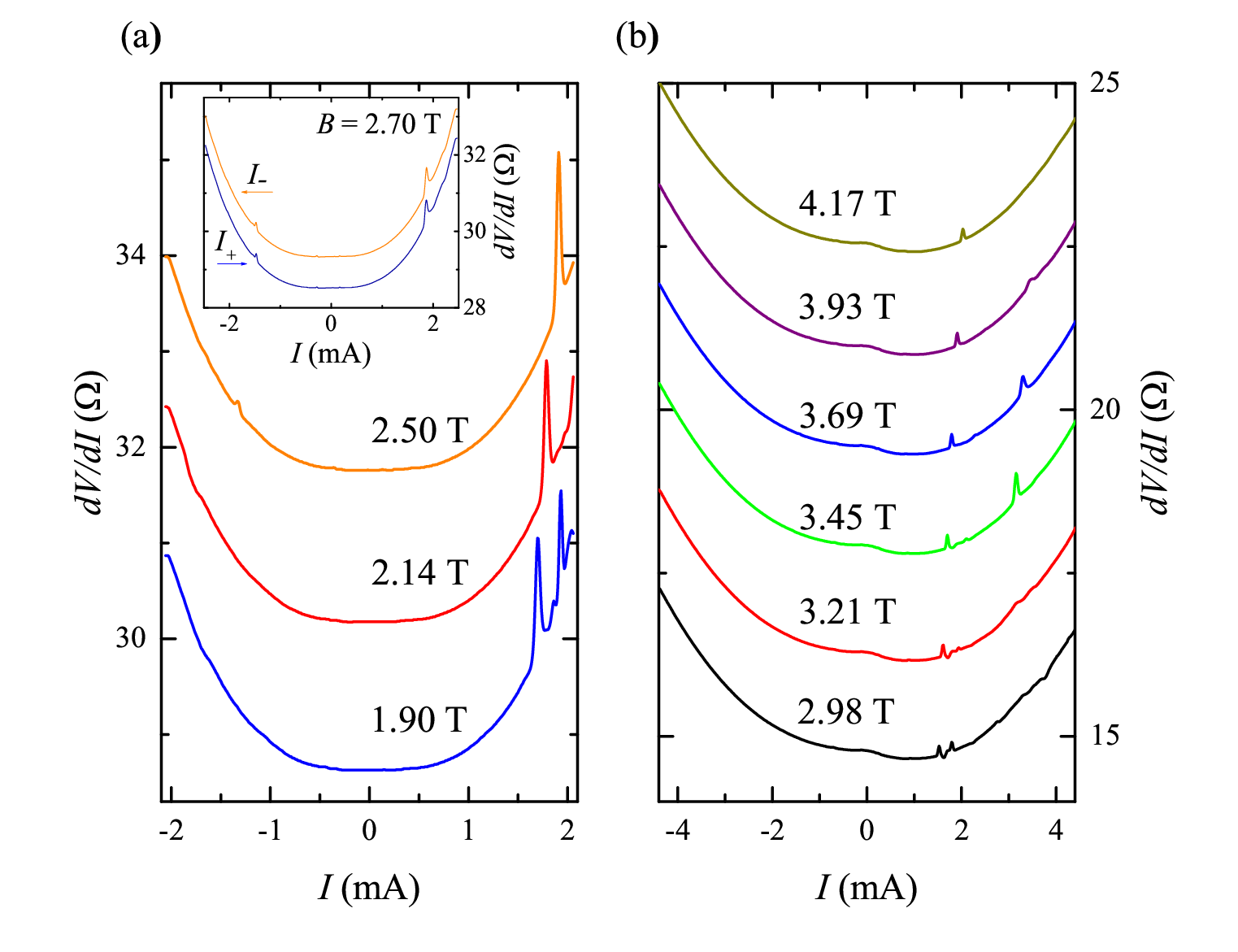}
\caption{Examples of $dV/dI(I)$ characteristics for two different samples, (a) and (b), respectively. The curves are shifted upward for clarity. $dV/dI$ is rising with both bias current polarities without any saturation at high biases. There are well-developed $dV/dI$ spikes for certain (positive) current polarity, their positions are sensitive to the magnetic field.   The inset to (a) demonstrates a lack of hysteresis for different current sweep directions. The magnetic field is perpendicular to the flake's plane. The measurements are performed at 4.2~K. }
\label{dVdI}
\end{figure}

Examples of $dV/dI(I)$ characteristics are shown in Fig.~\ref{dVdI} (a) and (b).  The curves demonstrate clear non-Ohmic behavior: $dV/dI$ is rising with both bias current polarities without any saturation at high biases. This behavior is  inconsistent with trivial  reasons like tunneling through different sample defects, which  generally leads to decreasing $dV/dI(I)$. Moreover, the experimental $dV/dI(I)$ curves are well reproducible for different samples with very different zero-bias resistance, cp. Fig.~\ref{dVdI} (a) and (b).  On the other hand, an overall symmetric increase in $dV/dI$ is usually attributed to electron scattering  in the magnetically ordered materials~\cite{concave_shape,myers}.

For the curves in Fig.~\ref{dVdI} we observe well-developed $dV/dI$ spikes for certain (positive) current polarity. These $dV/dI$ features are well reproducible in different cooling cycles. The spike position is sensitive to the magnetic field, as depicted in Fig.~\ref{dVdI}, it moves to higher currents  with increasing the magnetic field. In the same time, there is no noticeable hysteresis with the current sweep direction , see the inset to Fig.~\ref{dVdI} (a).  This behavior is very similar to one  reported for ferromagnetic multilayers~\cite{myers,tsoi1,tsoi2,katine,single,balkashin,balashov}, where the $dV/dI$ spikes have been attributed to spin-wave excitation modes. 

In addition, small spikes are sometimes present at the counter (negative) current polarity, as it is depicted in Fig.~\ref{dVdI} (a) for $B$ = 2.5~T and for $B$ = 2.7~T in the inset. These minor spikes have qualitatively the same magnetic field behavior,  so we  regard only the major ones at positive currents below.

\begin{figure}
    \includegraphics[width=\columnwidth]{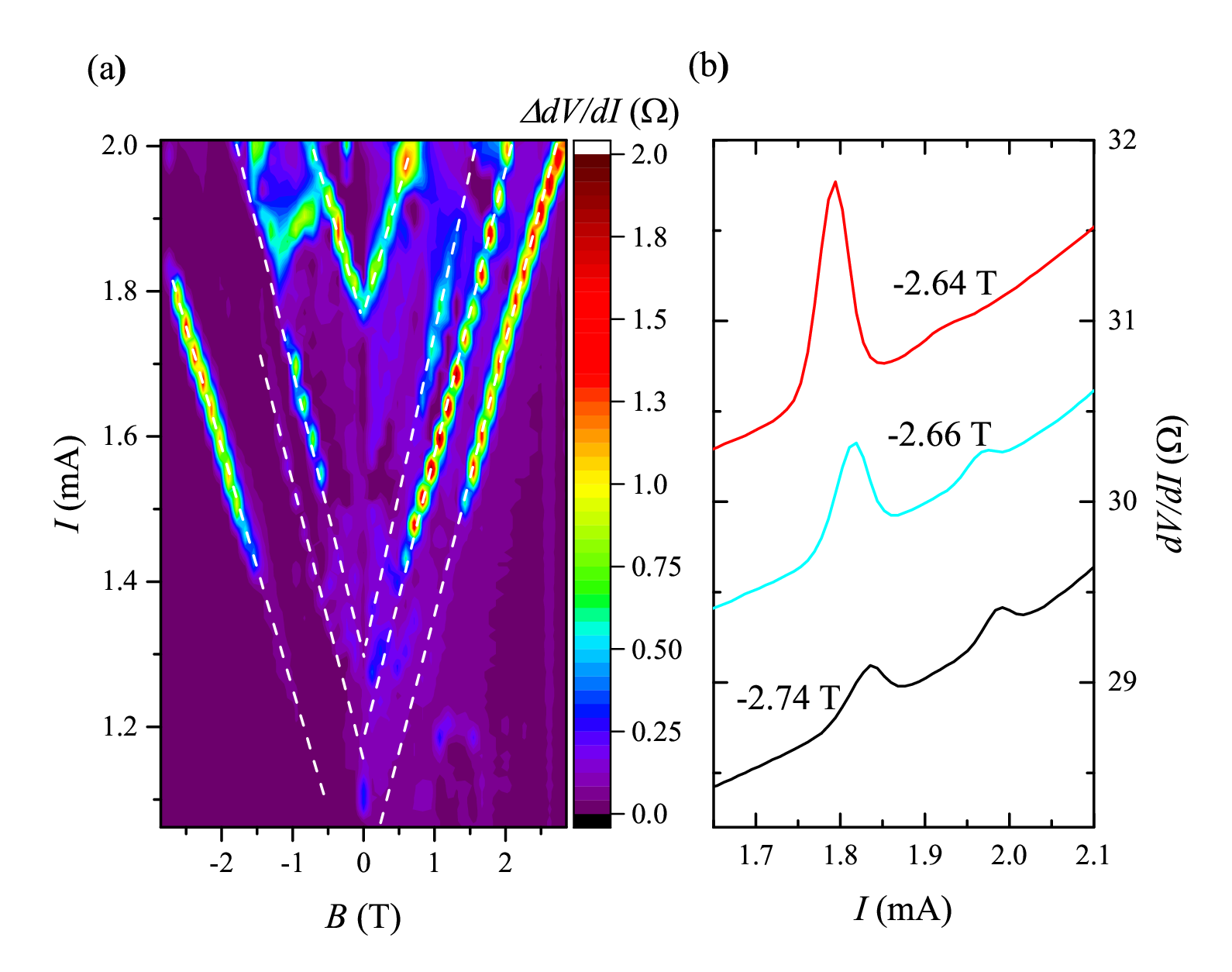}
    \caption{(a) Evolution  of $dV/dI$ spikes' positions  with magnetic field. The monotonous $dV/dI(I)$  is subtracted from the data, the colors correspond only to the spikes' amplitudes. There are several excitation branches with the linear field dependence, which are highlighted by  white dashed lines. The slopes coincide within experimental accuracy, but the spikes  vanish and reappear in some regions of the map. (b) An example of splitting of a particular branch in high magnetic fields, the curves are shifted downward for clarity.  While the amplitude of one spike is decreasing, the other spike is emerging at higher current, see also the corresponding region in (a).}
  \label{color}
 \end{figure}

The colormap in Fig.~\ref{color}~(a) demonstrates  evolution  of $dV/dI$ spikes  with magnetic field, which is applied normally to the Co$_3$Sn$_2$S$_2$ flake. Since the general $dV/dI(I)$ shape is practically independent of the magnetic field, see Fig.~\ref{dVdI},  it is subtracted from the data. Thus, the colors correspond only to the spikes' amplitudes in Fig.~\ref{color}~(a). 

It is clear, that the $dV/dI$ spikes' positions depend linearly on the magnetic field, as it has been also demonstrated~\cite{myers,tsoi1,tsoi2,katine,single,balkashin,balashov} for magnons~\cite{katine} in multilayers. In contrast to multilayers, we observe several excitation branches, which are highlighted by  white dashed lines on the colormap. The line slopes coincide within experimental accuracy, they also independent of the field direction.  The spikes are better visible at highest currents, where they can be demonstrated even in zero magnetic field.

On the other hand,  the relative amplitudes of the spikes  are different for different  branches, the spikes even vanish and reappear in some regions of the map.  Also, we observe splitting of a particular branch in high magnetic fields, see also  Fig.~\ref{color}~(b): while increasing the magnetic field, the spike's amplitude is diminishing, an additional spike appears at different current position; in higher fields,  two spikes coexist, see Fig.~\ref{color} (b). Such interplay generally reflects the ground state reconstruction~\cite{eisen,caf,smet}.

\section{Discussion}

$dV/dI(I)$ curves with an overall symmetric increase in $dV/dI$ and asymmetric $dV/dI$ spikes  have been reported for  spin valves~\cite{myers,tsoi1,tsoi2,katine,single,balkashin,balashov}. The spin valves are the sandwich structures, where spin-dependent scattering affects the magnetic moments of two spin-polarized layers, while their mutual orientation defines the differential resistance. We observe qualitatively similar effects in bulk properties of single-crystal Co$_3$Sn$_2$S$_2$ flakes. Since Co$_3$Sn$_2$S$_2$ is a quasi-two-dimensional ferromagnet, it can be regarded as a layered structure consisting of multiple Co monolayers. Thus, possible spin-transfer effects in a single Co layer should be  discussed firstly.

For spin valves,   the hysteresis in $dV/dI(I)$ curves in low magnetic fields indicates current-induced switching in the orientation of layers' magnetic moments between parallel and antiparallel orientations. In the high-field regime, the high-bias $dV/dI$ spikes correspond to a precessing spin-wave state with increasing deviations from the parallel alignment, which leads to increases in $dV/dI$ resistance.  In experiments on multilayers~\cite{myers,tsoi1,tsoi2,katine,single,balkashin,balashov},  spin-waves are excited by  injection of extremely high (~10$^9$ A/cm$^2$) current  density for thin, 2-4~nm, Co layers.

We estimate  the maximum current density as $\approx 10^4$ A/cm$^2$ for $I = 1$~mA through $S \sim 1\times 10 \mu$m$^2$ contact area, since the current density is maximal near the contact. Below, we rely on the model of Slonczewski~\cite{slonczewski}, which  was successfully adapted for quantitative analysis of spin-transfer effects in Co/Cu/Co nanopillars~\cite{katine}.  All the model parameters are well known for Co~\cite{katine}. 

Within the model~\cite{slonczewski,katine}, the $dV/dI$ spikes  positions $I_{sw}$ are described by  
\begin{equation}
I_{sw}(H) = \alpha \gamma e \sigma [H + H_{an} - H_{ex} + 2\pi M]/g(0), \label{eq}
\end{equation}
where $\alpha$ is the damping parameter, $\gamma$ is the gyromagnetic ratio, $\sigma$ is the total spin of the free layer, $M$ is the magnetization, $H_{ex}$ and $H_{an}$ are the exchange and the anisotropy  fields, respectively. The total spin is $\sigma = MV/\gamma \hbar$, where $V = Sd$ is the free layer's volume. $g(0) \leq$ 0.25 is the scalar function, which depends  on the mutual orientation of the free and static layers' magnetization~\cite{slonczewski}. The  $dI_{sw}/dH$ slope is defined by $\alpha \gamma e \sigma/g(0)$, so all the branches are parallel in Fig.~\ref{color}~(a).

A single Co monolayer plays the role of the free layer and the rest of the sample is the static layer. Then, one may estimate the free layer thickness  as $d \sim (dI_{sw}/dH) \cdot \hbar g(0)/\alpha e M S$. For known Co parameters~\cite{katine},  $\alpha$ = 0.005, $g(0)$ = 0.14, and $M$ can be estimated as  $\approx 100$~emu/cm$^3$ for Co$_3$Sn$_2$S$_2$. It gives inappropriately small $d \approx 3\cdot10^{-2}$~$\AA$. The damping parameter $\alpha$ can only be larger in a ferromagnet~\cite{tsoi1,tsoi2,single}, which leads to even smaller $d$. To obtain realistic $d \sim 4$~$\AA$, which is approximately equal to the distance between the Co layers in Co$_3$Sn$_2$S$_2$, one should use very small $S \approx$ 100$\times$100~nm$^2$, which is impossible even for inhomogeneous large 1$\times$10~$\mu$m$^2$ planar contacts. We wish to note, that $dV/dI$ spikes  are observed for different samples and are independent of the particular contacts, see Fig.~\ref{dVdI} (a) and (b).

Thus, possible spin-transfer effects between Co layers  can not explain the experimental results. The bulk of the sample is fully spin-polarized~\cite{kagome} above $\approx 0.5$~T, see Fig.~\ref{hall}, so  we can not attribute $dV/dI$ spikes to any bulk spin structures like domains. The domains are always accompanied by  hysteresis in $dV/dI(I)$ curves~\cite{myers,katine}, which can not be seen in the inset to Fig.~\ref{dVdI} (a). We have to search for Weyl specifics of the Co$_3$Sn$_2$S$_2$ flake. In principle, there are two possibilities, which refer to surface or bulk WSM properties, respectively:

(i) Broken time-reversal symmetry WSMs are characterized by topologically protected Fermi-arc surface states, which survive in high magnetic fields~\cite{armitage}. Low current density $\approx 10^4$ A/cm$^2$ can reflect the small total spin $\sigma$ in the surface state, which is unreachable in traditional  ferromagnets. 

(ii)   An additional bulk magnon branch is predicted~\cite{weyl_magnon} in magnetic WSMs. Physically, this magnon excitation can be understood as a direct result of the coupling between two magnetic moments mediated by Weyl fermions. The branch is also described by the modified Landau-Lifshitz equation~\cite{weyl_magnon}, so $I_{sw}(H)$ is still linear~\cite{slonczewski,katine}.  However, the gapless nature of Weyl fermions leads to long-range correlation of this magnon excitation~\cite{weyl_magnon}, which results in the dramatic decrease~\cite{dirac_magnon} in the damping parameter $\alpha$.  Thus, observation of low current density $\approx 10^4$ A/cm$^2$ well correspond to the predicted~\cite{weyl_magnon} novel magnon mode in magnetic WSMs.

In both cases, spin-unpolarized current from Au leads is injected into the magnetically-ordered WSM, which creates spin precession, or, in other words,  spin-wave excitation. The process is somewhat similar to the tip experiments~\cite{tsoi1,tsoi2,single}, while the current density is much smaller in our case. In multilayers, the asymmetry of $dV/dI$ spikes reflects the intrinsic asymmetry of the spin diode. The real Co$_3$Sn$_2$S$_2$ flakes are also not symmetric, see Fig.~\ref{cosns_sample}~(a), which creates higher current density for one region of the sample and is responsible for the spikes asymmetry in  Fig.~\ref{dVdI}. The latter variant (ii) is based on the solid theoretical background~\cite{weyl_magnon,dirac_magnon,magnon-first} and seems to be more realistic. In the terms of Eq.~\ref{eq}, different modes in Fig.~\ref{color}~(a)  are characterized by different  fields $H_{ex}$ and $H_{an}$, due to the complicated band structure of the real Co$_3$Sn$_2$S$_2$ WSM~\cite{kagome_arcs}, which is reflected in different onset current at zero magnetic field. Moreover, band bifurcation has been predicted in magnetic field~\cite{weyl_magnon}, which can be an origin of the  $dV/dI$ spikes' interplay in Fig.~\ref{color}~(b).

\section{Conclusion}

As a conclusion, we experimentally investigate electron transport in kagome-lattice ferromagnet Co$_3$Sn$_2$S$_2$, which is regarded as a time-reversal symmetry broken Weyl semimetal candidate. We demonstrate $dV/dI(I)$ curves  with  pronounced asymmetric  $dV/dI$ spikes, similar to those attributed to current-induced spin-wave excitations in ferromagnetic multilayers. In contrast to multilayers, there are several  $dV/dI$ spikes' sequences at low, $\approx$10$^4$ A/cm$^2$, current densities for a thick single-crystal Co$_3$Sn$_2$S$_2$ flake in the regime of fully spin-polarized bulk.  We attribute $dV/dI$ spikes at low current densities to novel magnon branches in magnetic  Weyl semimetals, which  can be understood as a direct result of the coupling between two magnetic moments mediated by Weyl fermions. In this case, the observed splitting of the magnon branches in magnetic field  may reflect the predicted band bifurcation for magnetic Weyl semimetal. Presence of spin-transfer effects at low current densities in Co$_3$Sn$_2$S$_2$ makes the material attractive for applications in spintronics.

\acknowledgments
We wish to thank V.T. Dolgopolov and Yu.S. Barash for fruitful discussions, and S.V~Simonov for X-ray sample characterization.  We gratefully acknowledge financial support partially by the RFBR  (project No.~19-02-00203), RAS, and RF State task.

\end{document}